\documentclass[12pt]{article}
\def\be{\begin{equation}}
\def\ee{\end{equation}}
\def\la{\label}
\def\bea{\begin{eqnarray}}
\def\eea{\end{eqnarray}}
\def\non{\nonumber}
\def\ci{\cite}
\def\la{\label}
\def\bib{\bibitem}

\def\lm{\lambda}
\def\Lm{\Lambda}
\def\Lmc{\Lambda_c}
\def\le{\left}
\def\ri{\right}

\def\gm{\gamma}

\def\Omdec{\Omega_{DE c}}
\def\Omdmc{\Omega_{DM c }}
\def\Omdgc{\Omega_{DG c}}

\def\Omde{\Omega_{DE}}
\def\Omdg{\Omega_{DG}}
\def\Omdeo{\Omega_{DE o}}
\def\Omdm{\Omega_{DM }}
\def\Omdmo{\Omega_{DM o}}

\def\Om{\Omega}

\def\Omro{\Omega_{r o}}
\def\Omrc{\Omega_{r c}}

\def\rdm{\rho_{DM}}

\def\rdmo{\rho_{DM o}}
\def\r{\rho}

\def\wdeo{w_{DE o}}

\def\wpo{w_{\phi o}}

\def\s8{\sigma_8}

\def\fr{\frac}

\usepackage[latin1]{inputenc}
\usepackage{graphicx}

\begin{document}


\begin{center}
   {\Large \bf  Dark Group: Dark Energy and Dark Matter }

\end{center}

\vspace*{0.5cm}

\begin{center}
{\bf A. de la Macorra\footnote{e-mail: macorra@fisica.unam.mx}}

\end{center}


\begin{center}
{\small
\begin{tabular}{c}
 Instituto de F\'{\i}sica, UNAM\\
Apdo. Postal 20-364, 01000  M\'exico D.F., M\'exico\\
\\
\end{tabular}
}
\end{center}


\begin{center}
{\bf ABSTRACT}
\end{center}
\small{ We study the  possibility that a dark group, a gauge group
with particles interacting  with the standard model particles only
via gravity, is responsible for containing  the dark energy and
dark matter required by present day observations. We show that it
is indeed possible and we determine the constrains for the dark
group.

    The non-perturbative effects  generated by a  strong gauge
coupling constant can de determined and a inverse power law
 scalar potential IPL  for the
dark meson fields is generated parameterizing the dark energy. On
the other hand it is the massive particles, e.g. dark baryons, of
the dark gauge group that give the corresponding dark matter. The
mass of the dark particles is of the order of the condensation
scale $\Lambda_c$ and the temperature is   smaller then the
photon's temperature. The dark matter is of the warm matter type.
The only parameters of the model are the number of particles of
the dark group. The allowed values of the different parameters are
severely restricted.   The dark group energy density at
$\Lambda_c$ must be $  \Omdgc \leq 0.17$ and the evolution and
acceptable values of dark matter and dark energy leads to a
constrain of $\Lmc$ and the IPL parameter $n$ giving
 $\Lambda_c=O(1-10^3)\,eV$ and  $0.28 \leq n \leq 1.04$.
 }

\vspace*{.1 cm}

\noindent \rule[.1in]{14.5cm}{.002in}

\thispagestyle{empty}

\setcounter{page}{0} \vfill\eject

\section*{Introduction}

The evidence for dark energy "DE" and dark matter "DM"  has been
established in the last few years. The measurements of high
redshift supernovae \ci{SN1a} show that the  universe is expanding
in an accelerating way requiring  an energy density with negative
pressure. In conjunction with the CMB spectrum \ci{map} and the
study of structure formation \ci{structure} show that the universe
is flat and with a matter content $\Om_b\simeq 0.05$ (baryonic),
 $\Om_{DM}=0.25\pm 0.1$ and a DE  $\Omde=0.7\pm 0.1$
 and an equation of state parameter   $\wdeo <-0.78$
(from now on the subscript o
represents present day quantities). Recent analysis
show that is must be smaller and closer to $-0.9$ \ci{neww}.

The physical process that gives rise to  dark matter and  dark energy
is yet unclear. Here we study the possibility that these two kinds
of energies which make $95\%$ of the universe are originated from
the same physical process. This connection allows for
a deeper insight into the nature of DE and DM.

The restrictions on DM is that it must account for $\Omdm=0.25\pm
0.1$  and it
should allow for structure formation at scales larger than  Mpc.
As we will see later our  models have a warm DM with a mass of the
order of $keV$. There are still problems with cold and warm DM
models. Cold DM have an overproduction of substructure of galactic
halos which  a warm DM model does not  have \ci{wdm}. On the other
hand, recent observations on the reionization redshift \ci{map}
seem to indicate that warm dark matter is not a good candidate.
However, the value of the parameters used are still not well established
which makes the conclusion not definite \ci{silk}. So, we believe
that further studies need to be done in order to fully set the
nature of dark matter.

The DE is  probably best described in terms of scalar fields or
quintessence. The possibility of having the quintessence field
parameterized by the condensate  of a gauge group has been studied
in \ci{bine,chris1,ax.mod}. Here, we want to study the possibility
that a gauge group contains at the same time the field responsible
for giving an accelerated universe at present time, i.e
quintessence, and on the other hand it gives the necessary amount
of DM needed for structure formation.

The starting point is a dark gauge group "DG"
 whose particles interact with the standard model "SM" only via gravity.
The requirement on the gauge group is that its gauge coupling
constant becomes strong at lower energies. When the coupling
becomes strong it will bind  the elementary dark fields together
at the phase transition or condensation scale $\Lmc$ (from now on
the subscript-c stands  for quantities defined at the condensation
scale $\Lmc$). Above this scale the particles are massless and at
the condensation  scale $\Lmc$ they acquire a mass of the order of
$\Lmc$. The elementary fields will form gauge invariant particles
due to the strong coupling.  These particles are dark
"mesons" and dark "baryons". The dark mesons  acquire a
non-trivial scalar potential $V$  below  $\Lmc$ and give rise to the DE
or quintessence field. Using Affleck-Dine-Seiberg superpotential,
which receives no quantum corrections, the scalar potential
takes the form of an inverse power law potential $V\sim \phi^{-n}$
with $n$ the inverse power law "IPL" parameter.
Besides the scalar field responsible for
quintessence we will have massive stable matter fields  and its
precisely these fields that account for the DM.   The appearance
of the quintessence field is only below the phase transition scale
and a late time accelerating epoch can be understood as the
consequence of having a small $\Lmc$.

We can further constrain the gauge group  by requiring that its
gauge coupling is unified with the standard model gauge couplings
\ci{chris1,ax.mod}. As an example we present our preferred model
\ci{chris1,ax.mod} which has an $SU(3)$ gauge group with $N_f=6$
chiral + antichiral fields with $g_{DG}=97.5$ degrees of freedom,
a condensation scale $\Lmc=42 eV$ and an inverse power potential
with $V=\Lmc^{4+n}\phi^{-n}$, $n=2/3$. This model gives a warm DM
with a free streaming scale $\lm_{fs}\leq  0.6 \, Mpc$ and an
equation of state parameter for the DE $w_{DE o}=-0.9$ with
$\Om_m=0.27, \Omdeo=0.73$.

\section*{Initial Conditions}

Let us now determine  the condition on the initial (i.e. at
$\Lmc$) dark energy  densities for DM and DE. Just before the
phase transition scale all dark particles are massless and  the
energy density of the DG can be written as $
\rho_{DG}=\fr{\pi^2}{30}g_{DG}T_D^4 $ where $T_D$  is the
temperature of the dark particles and in general it will be
different than the photon temperature $T_\gm$. The degrees of
freedom for a supersymmetric gauge group with $SU(N_c)$ and $N_f$
chiral plus antichiral fields is simply given by
$g_{DG}=(1+7/8)(2(N_c^2-1)+2N_fN_c)$ (the 7/8 count the fermionic
while 1 the bosonic degrees of freedom).
 After the phase transition we will
assume that the DE and DM have
$g_{DE}, g_{DM}$ degrees of freedom, respectively, with
$g_{DE}+g_{DM}=g_{DG}$ and $ \rho_{DG}=\rho_{DE} +\rho_{DM} $
where $\rho_{DE}=(\pi^2/30)g_{DE}T_D^4,\;\; \rho_{DM}=(\pi^2/30)g_{DM}T_D^4$
are the energy density
of the DE   and DM, respectively. At the condensation scale we can easily
 estimate the fraction
of the energy density for DM or DE in terms of the energy density
of the dark gauge group and their respective degrees of freedom
giving $\Omdmc=(g_{DM c}/g_{DG c})\Om_{DG }=(g_{DM c }/g_{DE c
})\Om_{DE c}$ with $\Om_i=\r_i/\r_{tot} $ where $\r_{tot}$ is the
total energy density which includes all the SM particles. We are
also assuming that there is conservation of energy within the DG,
i.e. the energy before and after the phase transition in the DG is
the same \ci{ax.mod}. We find it convenient to express the DM a
and DE in terms of $\Om_{DG}$ and $g_{DM c}, g{DE c}$ because
it shows that $\Omdmc,\Omdec$ cannot be arbitrary  small (or large)
since $g_{DM c}, g{DE c}$ must take values between one and $g_{DG}$.
Furthermore, $g_{DM c}, g{DE c}$ have clear interpretation in terms
of particle physics.

The number of the relativistic degrees
of freedom is a time (energy) dependent quantity and therefore
we write a subscript $c$ on the degrees of freedom at the
condensation scale. At later times $g_{DM}$ may be smaller than
$g_{DM c}$ since dark matter particles will decay into the
lightest stable particles, the dark "LSP".

The standard model and the DG will in general not have the same
temperature, i.e. $T_{D}\neq T_\gm$. We can use $T_D$ and $T_\gm$
as independent variables or without loss of generality we
can express $T_D$ as a function of $T_\gm$ and $g_{dec}$ (we still
have two independent variables), the
number of degree of freedom of the SM at an energy scale where
$T_\gm=T_D$. However, we chose to take $T_\gm$ and $g_{dec}$ as the independent
variables since we can relate  $g_{dec}$ more directly to particle physics.

If the standard model and the DG have the same initial temperature
we can use entropy conservation to determine the  relative
temperature between the standard model $T_\gm$ (photon's
temperature) and the DG $T_{D}$ at a lower scale. Since  these
groups (SM and DG) interact via gravity only they  would  not
maintain a thermal equilibrium with each other. The same initial
temperature can be obtained if   the gauge groups are unified at
the unification scale $\Lm_{gut}=10^{16} GeV$ and/or if the
reheating process after inflation is gauge blind and gives the
same amount of energy to all relativistic degrees of freedom, the
democratic reheating. So, from entropy conservation we obtain the
relative temperature between the standard model $T_\gm$  and the
DG $T_{D}$  for relativistic degrees of freedom $T_{D}=T_\gm
(g_{sm f}g_{DG dec}/g_{sm dec}g_{DG f})^{1/3}$ where $g_{dec}\equiv
g_{sm dec}, g_{sm
f}, g_{DG dec}, g_{DG f}$ are the relativistic degrees of freedom
at a final stage and at decoupling scale (which is {\it not}
$\Lmc$) for  SM and DG, respectively. The energy ratio is
given by $\Om_{DG f}=g_{DG f}(T_{D}/ T_{\gm})^4/(g'_{sm f}+ g_{DG
f}(T_{D}/T_{\gm})^4) $ where $g'_{smf}$ takes into account all SM
relativistic degrees. For a neutrino
one has at  decoupling  $g_{dec}=11/2, g_{sm f}=2$ and with
$g_{DG f}=g_{DG dec}$ one has $T_\nu=T_\gm (4/11)^{1/3}=
(1/1.76)T_\gm$. However, if the decoupling is at a high energy
scale, say  $T\gg 10^3 GeV$, then all particles of the standard
model are still relativistic and
$T_{D}=T_\gm (43/11/g_{dec})^{1/3}$
for $T_\gm < 1\, MeV$ ($g_{sm f}=43/11$ takes into account
neutrino decoupling). We get a temperature  $T_D \simeq  (1/3)T_\gm$
for the SM with $g_{dec}=106.75$ and $T_{D}\simeq
(1/3.88)T_\gm$ for the minimal supersymmetric SM (MSSM) with
$g_{dec}=228.75$.
 The temperature of DG is in these
cases $3-4$ times smaller then the photon temperature and $2-3$
times smaller then $T_\nu$. If there are more relativistic
degrees of freedom coupled to the susy-SM (could be Kaluza-Klein
states or other gauge  groups, e.g.
gauge group responsible for susy breaking \ci{ax.mod})
at decoupling then $T_D$ would be even smaller.

\subsection*{Limits on the Gauge Group Degrees of Freedom}

We can set an upper and lower limit to $\Omdg$. The smallest
number of degrees of freedom would be for a gauge group with
$N_c=2, N_f=1$ giving $g_{DG}=18.75$. While the upper limit on
$g_{DG}$ comes from Nucleosynthesis "NS" bounds which requires an
upper limit to any extra energy density. This limit is
$\Omdg(NS) \leq 0.1-0.2$ \ci{NS}. Since
 $g_{DG}/g_{dec}^{4/3}=
(10.75)^{-1/3}\Omdg(NS)/(1-\Omdg(NS))$ the NS bound sets un upper
limit $g_{DG}\leq 0.05g_{dec}^{4/3},\,0.113 g_{dec}^{4/3}$ for
$\Omdg(NS)\leq 0.1,0.2$, respectively. Taking $  g_{DG} \leq 0.113
g_{dec}^{4/3} $ we obtain un upper limit $\Omdgc \leq 0.17$ at any
condensation scale $\Lmc$ below $1 MeV$. We have $\Omdgc <
\Omdg(NS)$ for $\Lmc<1\,MeV$ due to neutrino decoupling and the
electron and positron acquiring a mass.

\section*{Dark Matter}

\subsection*{Free Streaming Scale}

Before studying the dynamics of the DG let us determine
the constraint on the temperature and mass for DM
in order to agree with structure formation. The free streaming
scale $\lm_{fs}$ gives the minimum size at which perturbations
survive. For scales smaller than the $\lm_{fs}$ the perturbations
are wiped out. For structure formation it is required
that $\lm_{fs} \leq  O(1) Mpc$.
One has \ci{lfs}
\bea\la{free}
\lm_{fs}&\simeq& 0.2 (\Omdm h^2)^{1/3}(1.5/g'_{DM})^{1/3}
(keV/m)^{4/3}\\
&=&  0.079 (\Omdm h^2)^{-1}(g'_{DM}/1.5)
(228/g_{dec})^{4/3}\non
\eea
where $g'_{DM}=g_{b DM}+3/4 g_{f DM}$
with $g_{b DM}$ the bosonic,
$g_{bf DM}$ the fermion degrees of freedom of DM, i.e.
the LSP,
and we  used eq.(\ref{omdm1}) in the second equality of
eq.(\ref{free}).

\subsection*{Constraint on the Mass of the Dark Matter Particle }

Lets us now study the constraint on the mass of the LSP.
The energy density of the DG
will be divided in  DE (quintessence) and
 DM. For DM the entropy
conservation gives $n_{DM}/n_{\gm}=(g'_{DM}/2)(T_{D}/T_\gm)^3 $
where $n_{DM}, n_\gm=2(\zeta(3)/\pi^2)T_\gm^3$ are the number density for
DM and photon respectively. Since the energy
density for matter is $\rho_m=nm$ and using
$ \rho_\gm=n_\gm (\pi^4/30\zeta(3)) T_\gm$ we can write
$\Om_{DM o} =\Om_{\gm o}(\zeta(3) 30/\pi^4)
(n_{DM}/n_\gm)(m/T_{\gm o})=
\Om_{\gm o}(\zeta(3) 30/\pi^4)(g'_{DM}/2)(m/T_{\gm o})(T_D/T_\gm)^3$
 giving \be
\Om_{DM o}=0.25\le(\fr{0.71}{h_o}\ri)^2
\le(\fr{g'_{DM}m}{g_{dec}1.66\,eV}\ri) \la{omdm1} \ee where we
have used in the last equation the present day quantities
$h_o^2\Om_\gm=2.47\times 10^{-5}$, $T_{\gm o}=2.37\times 10^{-13}
GeV$. Eq.(\ref{omdm1}) is valid  for all DM that decouples at
temperature $T_i \gg 10^3 GeV$ from the susy-SM. Taking the
central values of wmap \ci{map} $\Omdmo h^2_o=0.135-0.0224=0.1126$
(where $\Omega_bh^2=0.0224$) one gets a neutrino mass $m=12\,eV$
and  $\lm_{fs}=36\,Mpc$ giving the usual hot DM problem. It cannot
form structure at small scales. For a model that decouples from
the susy-SM at $T\gg 10^3 GeV$ one has $T_{D}/T_\gm \leq 1/3.88$
with $ g_{dec}\geq 228$, a mass $m \geq 254 \, eV$ for $g'_{DM}=
1.5$ (i.e. a fermion)  and eq.(\ref{free}) gives $\lm_{fs}\simeq
0.41\,Mpc$. Allowing for a more conservative variation of $\Om_{DM
o}=0.25\pm 0.1$ and $h_o=0.7\pm 0.05$ the constraint on
$g'_{DM}\,m/g_{dec}$ from eq.(\ref{omdm1}) is $0.83 g_{dec} eV\leq
g'_{DM}\,m\leq 2.59 g_{dec} eV$. The number of degrees of freedom
$g'_{DM}$ is not arbitrary since $0.113 g_{dec}^{4/3} \geq g_{DG}
> g'_{DM} \geq 1$, as discussed above. This bound implies that the
mass of the DM particle must  be
 \be\la{mass} 1.2
(228/g_{dec})^{1/3}\,eV \leq m \leq 593 (g_{dec}/228)\,eV.
\ee
For $g_{dec}\leq 228$ we have  $m\leq 593 \,eV$ and we would get a larger
mass $m\geq 750\,eV, 1 keV$ if  $g_{dec}\geq 675,
900$ for $g'_{DM}=1.5$.

\subsection*{Constraint on the Condensation Scale $\Lmc$ and on the
IPL parameter $n$}

In order to connect the dynamics of the dark energy (quintessence)
and the constraint on dark matter density we
evolve the  DM from present day to the phase transition
scale $\Lmc$ where the particles acquired a mass.

The evolution of
the DM is   $ \rdmo=\rdm(a/a_o)^{3}$ where $a(t)$ is the scale
factor. In terms of $\Omdm=3H^2\rdm$ (we have taken $8\pi
G=1/m^2_{pl}=1$) we can write the DM energy density as
\be\la{omdmi}
\Omdmo=\Omdmc (\Omro/\Omrc)^{\fr{3}{4}}
(H_c^2/H_o^2)^{\fr{1}{4}} \ee
where we have expressed the scale factor $a$ in terms of the
relativistic energy densities,
$a_c/a_o=(\Omro H^2_o/\Omrc H^2_c)^{1/4}$.
The evolution of the DE depends on
the specific potential. However, the non-abelian gauge dynamics
leads to an inverse power potential of the form \ci{bine,chris1,ax.mod}
 \be\la{v}
V=\Lmc^{4+n}\phi^{-n} \ee where $\phi=<\bar Q Q>$ is the
condensate of the elementary fields. Here we will treat $n$ as a
free parameter but it can be related to $N_c,N_f$ by
$n=2+4\nu/(N_c-N_f)$ and $\nu$ counts the number of light
condensates \ci{chris1,ax.mod}. When the kinetic term is much
smaller than the potential energy one has $\Omde \simeq
\Lmc^{4+n}\phi^{-n}/3H^2$. This is certainly valid for present day
since we require $\rho_{DE}$ to accelerate the universe and the
slow roll condition $E_k \ll V$ must be satisfied. Since  the
beginning of an accelerated epoch is very recently one has
$\phi_o\simeq 1$ \ci{bine}. Of course, a numerical analysis must
be performed \ci{chris1,ax.mod} in order to obtain the precise
values of $\phi_o, \wpo$ but the analytic solution is a reasonable
approximation. At the condensation scale $\Lmc$ the initial value
of the condensate $\phi_c$ must be giving by $\Lmc$ and taking
$\phi_c=\Lmc$ \ci{chris1}  we have
\be\la{lmc}
\Omdec=\fr{\Lmc^4}{3H_c^2},\hspace{1cm}
\Omdeo=\fr{\Lmc^{n+4}}{3H_o^2}. \ee Using eqs.(\ref{omdmi}) and
(\ref{lmc}) we can write
\be
\la{odmde} \Omdmo= \Omdmc
(\Omro/\Omrc)^{\fr{3}{4}}
(\Omdeo/\Omdec)^{\fr{1}{4}}\Lmc^{-\fr{n}{4}}
\ee
where we have
used  $H^2_o/H_c^2=(\Omdec/\Omdeo)\Lmc^n$. An easy estimate
of the order of magnitude for
$\Lmc$ and $n$ can be obtained from eqs.(\ref{lmc})  and eq.(\ref{odmde})
using
$\Omdmo=0.25,\Omdeo=0.7,\Omro=O(10^{-5})$
and $\Omdmc=O(10^{-2}),\Omdec=O(10^{-1}),\, \Omrc=O(1)$
giving
\bea\la{lmc2a}
\Lmc&\simeq& H_o^{2/(4+n)}\simeq 10^{-120/(4+n)}\\
&\simeq & 10^{-20/n}
\la{lmc2b}\eea
where we have used $H_o\simeq 10^{-60}$ (in Planck units).   From
eqs.(\ref{lmc2a}) and (\ref{lmc2b}) we get a rough solution
for  the IPL parameter $n$ and condensation scale,
\be
n\simeq 4/5, \hspace{1.5cm} \Lmc\simeq 200\,eV.
\ee
In general eq.(\ref{odmde}) depends also
on $g_{dec}, g_{DG}, g_{DM c}$ through $\Omdmc,\Omdec$.
Taking as a concrete  example a dark gauge group with
$g_{DG}=97$  with $g_{DM c}=1.5$ and the
central wmap values $h_o=0.71, $ $ \Omdmo =0.25$ \ci{map} we find
from eqs.(\ref{lmc})  and eq.(\ref{odmde})
for the MSSM $g_{dec}=228$ and for $g_{dec}=900$
an inverse power parameter $n=0.78,\, 0.9$
and $\Lm_c=189,\,745\,eV$,  respectively.

We can determine the
allowed range of values of $n$ and $\Lmc$, which  is
quite limited, if
 we allow for a more conservative variation
 $\Omdmo=0.25\pm 0.1$, $h_o=0.7\pm 0.05$,
   $228\leq g_{dec}\leq 900$
 (the upper value gives a dark mass of $m\simeq 1\,keV$ (see below
 eq.(\ref{mass}))) and with $1\leq g_{DM c}\leq g_{DG c}-1$.
The range for $n$ and $\Lm_c$ if we take a dark gauge group with  $g_{DG c}=97$
and the largest gauge group allowed by NS $g_{DG c}=0.113 g_{dec}^{4/3}$
(results in parenthesis) is shown
in table 1  for different values of $g_{dec}$,
the MSSM $g_{dec}= 228$, $g_{dec}=675$ (this value gives
a dark mass $m\simeq 750\,eV$)  and $g_{dec}= 900$. In all cases  the lowest limit is given by $  g_{DM c}=g_{DG
c}-1, h_o=0.65, \Omdmo=0.15$ while the upper limit has  $g_{DM
c}=1, h_o=0.75, \Omdmo=0.35$.
\begin{center}
\begin{table}
\begin{center}
\begin{tabular}{|c|c|c|c|c|}\hline
   $g_{dec}$ & $n_{min}$ & $n_{max}$ & $\Lm_{c\,min}eV$ & $\Lm_{c\,max}eV$ \\ \hline
   228 & 0.34 (0.31) & 0.87 (0.88)& 0.55 (0.34) & 518 (585)\\
   675 & 0.42 (0.29)& 0.96 (1.0)  & 1.63 (0.23) & 1530 (2484)\\
   900 & 0.44 (0.28) & 0.98 (1.04) & 2.17 (0.21) & 2040 (3639)\\ \hline
 \end{tabular}
\end{center}
\caption{We show the minimum and maximum values of $n$ and $\Lmc$
for different $g_{dec}$ and for a gauge group with $g_{DG}=97$ and
$0.113g_{dec}^{4/3}$ (results in parenthesis). The lowest limit
has $g_{DM c}=g_{DG
c}-1, h_o=0.65, \Omdmo=0.15$ while the upper limit has  $g_{DM
c}=1, h_o=0.75, \Omdmo=0.35$.}
 \end{table}
\end{center}
From table 1 we see that the allowed range is
\be\la{cons} 0.28 \leq n \leq
1.04 \;\;\Leftrightarrow\;\; 0.21 \,eV \leq \Lm_c \leq 3639\, eV
\ee
valid for   $g_{dec}\leq 900$. Increasing $g_{dec}$ would
enlarge the range of $n, \Lm_c$ but not significantly,
the variation in $n$ from $g_{dec}=228$ to 900 (i.e. almost
$400\%$) increases the upper value of $n$ by
$13\%$ while it has a linear effect on $\Lmc$.
In fig.1 we
show the behavior of $\Omdmo$ as a function of $n$ for different
values of $g_{DG c} =0.113g_{dec}^{4/3}$ with
$g_{dec}=228,675,900$ (solid,dashed and dotted lines,
respectively) for the extreme values of $g_{DM c}$ given  by
$g_{DG}-1 \geq g_{DM c}\geq 1$. The allowed region is in between
the horizontal lines $\Omdmo=0.15-0.35$. From eq.(\ref{cons})  we
see that there is only a limited range of condensation energy
scales and IPL parameter $n$ that allows for a gauge group to give
the correct DM and DE densities. It is also interesting to note
that the lower limit on $\Lmc$ is very similar to the one obtain
by CMB analysis \ci{ax.mod} where the minimum scale was  $\Lmc=0.2
\,eV$. On the other hand, the evolution of quintessence requires
for $\Omdec < 0.17$ an IPL parameter $n$ to be smaller than $n
\leq 1.6
 $ for $w_{DE o} \leq -0.78$ which is the upper value of wmap.
 For smaller $\Omdec$ we will need a smaller $n$, e.g.
 $\Omdec =0.05$ requires $n\leq 1.05$. So,
once again there is a consistency within the acceptable values of
$n$ coming form different considerations (amount of DM and
observable $w_{DE o}$).
\begin{figure}[ht!]
\begin{center}
\includegraphics[width=10cm]{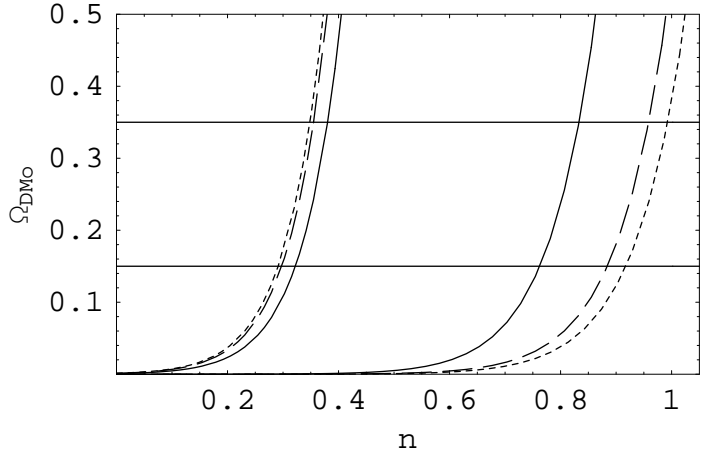}
\caption{\small{We plot $\Omdmo$ as a function of the IPL
parameter $n$. The allowed region is the one between the
horizontal lines $\Omdmo=0.15-0.35$ and the curves with the
limiting values of $1.5\leq g_{DM}\leq g_{DG}-1$ for
$g_{dec}=228,675,900$ (solid,dashed and dotted lines,
respectively). }} \label{fig1}
\end{center}
\end{figure}
The constraint on $\Lmc$ is very similar
to the constraint obtained for the DM particle mass $m$ obtained
in eq.(\ref{mass}). The similarity  $m\sim \Lmc$
is required by non-abelian gauge dynamics and it
is indeed satisfied as can be verified
using eqs.(\ref{omdm1}), (\ref{lmc}) and (\ref{odmde})
\be\la{mLm}
c\equiv\frac{m}{\Lm_c}=\fr{\pi^4}{\zeta(3) 30}\; \fr{g_{DM c}}
{g'_{DM}}
\ee
with $\pi^4/(\zeta(3) 30)\simeq 2.7$.
There is a subtle
point on the values of $g_{DM c}, g'_{DM}$. The "true" degrees
of freedom of the dark matter particles (i.e. the lightest
field of the dark gauge group)  are given by $g'_{DM}$
while $g_{DM c}$ and $g_{DE c}$ represent
the proportion of energy density that goes
into $\Omdmc$ and $\Omdec$. It is   reasonable to assume
that the particles of the dark group will decay
into the lightest state. Therefore we expect $g_{DM c} > g'_{DM}$
and $m > \Lm_c$.

\section*{Dark Energy}

Having established the necessary condition on the initial DM we
will study the dark gauge group. The idea is based on the work
\ci{chris1,ax.mod} and details can be obtained there. Here we will
only sketch the arguments. If a gauge group has a gauge coupling
constant that increase with decreasing energy than the elementary
fields in the group will be bind together at the phase transition
scale $\Lmc$ where the coupling becomes strong. At strong coupling
the dynamics becomes non-perturbative and for non-abelian gauge
group we can use the superpotential of Affleck-Dine-Seiberg "ADS"
\ci{ADS} to determine the scalar potential generated at $\Lmc$.
This superpotential is exact since it receives no quantum corrections.
For $N_c>N_f$ the only gauge singlet fields that arise are dark
"mesons" in the form of $<\bar Q Q>$ (as meson fields in QCD). It
is this field (more precisely it is the lightest meson field) that
is the quintessence field $\phi$ and  gives the DE. The potential
generated by ADS is given by eq.(\ref{v}) \ci{bine,chris1,ax.mod}
$V(\phi)=\Lmc^{4+n}\phi^{-n} $ where the quintessence field is
$\phi^2=<\bar Q Q>$ and the parameter $n$ is $n=2+4\nu/(N_c-N_f)$,
where $N_c, N_f, \nu$ are the number of colors, chiral fields and
lightest meson field. It is reasonable to assume that at the
condensation scale
 $\phi_c=\Lmc$ since it is the relevant scale of the physical process.
If we have $N_c<N_f$ then on top of the gauge singlet meson
fields we can have gauge singlet dark baryons
$B^{i,...,N_c}=\prod_i^{N_c} Q^i$ and anti baryons. These
particles get a non-vanishing mass due to non-perturbative effects
(like protons and neutrons in QCD).  These baryons could be
degenerated in mass or there could be a lightest massive stable
baryon. The order of magnitude of the mass of the DM particle can
be estimated by the condensation
\be\la{mb}
m=c\,\Lmc
\ee with $c=O(1)$ a constant. Eq.(\ref{mb}) should be compared
with eq.(\ref{mLm}).
Eqs.(\ref{v}) and (\ref{mb}) set the cosmological evolution for DE
and DM. In this picture we have at high energies $E>\Lmc$ a DG,
i.e. a non-abelian gauge group that interacts with the standard
model only via gravity, with massless particles and redshifting as
radiation. At $\Lmc$ non-perturbative effects, due to a strong
coupling, generate a mass for dark baryons and a scalar potential
for dark meson. The DM is the massive  stable particle with mass
given by eq.(\ref{mb}) while the quintessence with potential
(\ref{v}) gives the DE. The free parameters of the models are
$n,\Lmc$ and the energy density at the condensation scale. All
these quantities  can  be determined in terms of the number of
degrees of freedom (i.e number of particles). Different values of
$n,\Lmc$  may lead to different acceptable models.

\section*{Conclusions}

Let us  conclude and summarize the main results. We have studied
the possibility that a dark gauge group contains the dark matter
and dark energy. The allowed values of the different parameters
are severely restricted by different considerations. The NS
constrain on $g_{DG}$ sets a limit to the dark energy density at
$\Lmc$ of $ \Omdgc \leq 0.17$. The evolution and acceptable values
of DM and DE leads to a constrain of  $\Lmc$ and  $n$ giving
 $0.21\,eV \leq \Lmc \leq 3639 \,eV$ and $0.28 \leq n \leq 1.04$
 for $g_{dec} \leq 900$. The mass of the dark particle would
 be of the order of $1-10^3 \,eV$,   depending on the value of $g_{dec}$,
 giving  a warm dark matter.
 For larger  values of $g_{dec}$ one gets a larger mass.

On the other hand, the analysis of the CMB spectrum sets also a
lower scale for the condensation scale $\Lmc > 0.2 \; eV$ with
$n>0.27$. The evolution of the quintessence field requires also a
small $n$ in order to have a small $\wdeo$. For $\Omde\leq 0.17$
and $ \wdeo \leq -0.78$ one needs $n < 1.6$. So, from
three different analysis (quintessence, DM
and CMB spectrum)  we are led to conclude that the most acceptable
models have a low condensation scale $\Lmc$ of the order of
$1-10^3\,eV$. The fact that the condensation is low explains why
the acceleration of the universe is at such a late time.

Acknowledgments

This work was supported in part by CONACYT project 32415-E and
DGAPA, UNAM project IN-110200.

\end{document}